\documentclass[sigconf, screen]{acmart}

\usepackage{svg}
\usepackage{booktabs}
\usepackage{tabularx}
\usepackage{graphicx}
\usepackage{placeins}
\usepackage{afterpage}

\usepackage{subcaption}   
\usepackage{cleveref}

\usepackage{fvextra}

\crefname{figure}{Figure}{Figures}
\Crefname{figure}{Figure}{Figures}

\usepackage[T1]{fontenc}
\usepackage[utf8]{inputenc}
\usepackage{cmap}
\usepackage{microtype}

\usepackage{newunicodechar}
\newunicodechar{–}{--}   
\newunicodechar{—}{---}  
\newunicodechar{’}{'}    
\newunicodechar{‘}{`}    
\newunicodechar{“}{``}   
\newunicodechar{”}{''}   

\setcopyright{none}
\copyrightyear{2026}
\acmYear{2026}
\acmDOI{XXXXXXX.XXXXXXX}
\acmConference[EASE 2026]{The 30th International Conference on Evaluation and Assessment in Software Engineering}{9–12 June, 2026}{Glasgow, Scotland, United Kingdom}

\acmISBN{}

\begin{document}

\title{Automated Classification of Human Code Review Comments with Large Language Models}


\author{Semih Çağlar}
\affiliation{%
  \institution{Bilkent University}
  \department{Computer Engineering} 
  \city{Ankara}
  \country{Turkey}}
\email{semih.caglar@ug.bilkent.edu.tr}
\orcid{0009-0003-1311-5747}

\author{Şükrü Eren Gökırmak}
\affiliation{%
 \institution{Bilkent University}
 \department{Computer Engineering} 
 \city{Ankara}
 \country{Turkey}}
\email{eren.gokirmak@ug.bilkent.edu.tr}
\orcid{0009-0003-2128-7872}

\author{Eray Tüzün}
\affiliation{%
  \institution{Bilkent University}
  \department{Computer Engineering} 
  \city{Ankara}
  \country{Turkey}}
\email{eraytuzun@cs.bilkent.edu.tr}
\orcid{0000-0002-5550-7816}

\renewcommand{\shortauthors}{Caglar et al.}

\begin{abstract}
\noindent
\textbf{Context:} 
Code reviews are essential for maintaining software quality, yet many human review comments suffer from issues such as redundancy, vagueness, or lack of constructiveness. These types of comments may slow down feedback and obscure important insights. Prior work on code review comments mostly explore the detection and categorization of useful comments, while fine-grained categorization of comment issues remains underexplored.

\noindent
\textbf{Objective:}
This work aims to design and evaluate an automated system for classifying code review comments according to specific categories of issues.

\noindent
\textbf{Methodology:}
We introduced a nine-label taxonomy for code review comments, covering six review comment smells and three common useful intents, and manually labeled 448 comments from a publicly available dataset. We benchmarked zero-shot and one-shot single-label classification over each comment and its associated unified diff hunk, comparing GPT-5-mini, LLaMA-3.3, and DeepSeek-R1. We reported macro-F1 as the primary metric.

\noindent
\textbf{Results:}
Zero-shot performance was moderate under class imbalance (macro-F1 0.360 to 0.374). One-shot exemplar conditioning had model-dependent effects: GPT-5-mini and DeepSeek-R1 macro-F1 scores improved, however LLaMA-3.3 suffered a slight decrease. Exemplars most consistently helped intent-boundary labels, whereas classification of evidence-sensitive labels remain challenging.

\noindent
\textbf{Conclusion:}
Our results indicate that comment--diff evidence is sufficient for some labels but limited for evidence-sensitive smells. Future work includes adding thread context, improving intent-preserving rewrites, and validating robustness across platforms.

\end{abstract}



%
\maketitle


\section{Introduction}
\label{sec:intro}



Code review (CR) is a collaborative process where developers asynchronously examine peers’ code changes to find defects and suggest quality improvements, while also facilitating knowledge transfer among developers~\cite{modern_code_review_bacchelli_intro,umutcihanpaper}. 
This practice evolved as a pragmatic alternative to the resource-intensive Fagan inspections introduced in the 1970s~\cite{fagan_inspection}. By reducing formal meeting overhead and enabling distributed, tool-supported collaboration, code review has become common practice across both commercial enterprises~\cite{modern_code_review_bacchelli_intro,rigby_bird_convergent_software_peer_review_intro,google_code_review_case_study} and open-source ecosystems~\cite{rigby_bird_convergent_software_peer_review_intro}.
Developers allocate roughly 10--15\% of their working time to performing code reviews~\cite{bosu2017_process_aspects_code_review}. Developers spend on average six hours per week on reviews, with a median turnaround of 24 hours for sign-off, and many reviews dragging on for days \cite{code_reviews_slows_us, DOGANCodeReviewSmells,Altun2025}. While beneficial, CR demands substantial human effort and time, often leading to backlogs and delays~\cite{umutcihanpaper}. Empirical studies at Microsoft report that only about 15\% of review comments indicate potential defects, while more than 50\% focus on maintainability concerns~\cite{code_reviews_slows_us}. 

The essential components of modern code review are the review comments~\cite{bosu2015useful}. Patch authors typically refer to these comments when revising their code modifications and preparing the subsequent patch submission. Therefore, the quality of these comments plays a crucial role in the effectiveness of the review process and must be sufficiently useful to guide meaningful code improvements. An earlier large-scale study found that roughly 34.5\% of review comments across five large Microsoft projects were non-useful~\cite{bosu2015useful}. We refer to such low-value comments as \emph{review comment smells}.
Consequently, improving code review effectiveness is a high priority for many organizations~\cite{bosu2015useful,hasan_balanced_scorecard_code_review_intro}. 

Most prior studies evaluate review comments along a single quality axis. Early work at Microsoft labeled 1.5 million comments as \emph{useful} vs.\ \emph{non-useful} and trained a decision-tree classifier to predict this dichotomy~\cite{bosu2015useful}. Subsequent detectors such as \emph{RevHelper} predict usefulness at submission time on an industrial corpus of 1{,}116 comments~\cite{predicting_usefulness_of_code_review_comments_rahman}, while other work distinguishes \emph{actionable} vs.\ \emph{non-actionable}~\cite{rahman2024}. Later studies, such as the \emph{EvaCRC} framework focuses on multiple axes of quality by grading comments on four attributes (emotion, question, evaluation, suggestion) mapped to quality tiers from poor to excellent, but it does not identify concrete smell types~\cite{EvaCRC}. Likewise, the \emph{ClearCRC} framework evaluates comments on three attributes (relevance, informativeness, and expression) but do not map them to concrete smell types \cite{chen_clear_crc}. Collectively, these studies show automatic measurement is feasible, but they do not explain \emph{why} a comment is low value.

A complementary line of work proposes topic-oriented taxonomies. Li et al.~\cite{zhixing} define four high-level categories (\emph{Correctness}, \emph{Decision}, \emph{Management}, \emph{Interaction}) with 11 subcategories, and Ochodek et al.~\cite{journal_classification} introduce CommentBERT, which covers 12 content themes (e.g., logic, API, documentation). These approaches primarily characterize the topics or functions of review comments and typically presuppose comments to be substantive and useful. While some taxonomies include categories that are adjacent to comment quality or potential smells, such categories are generally coarse-grained and do not distinguish between concrete types of quality issues. In contrast, our work focuses explicitly on categorizing low-quality review comments into fine-grained smell categories. To our knowledge, no operational taxonomy explicitly targets review comment smells at this level of granularity.

Parallel to the work on review comment classification, automated review comment generation has gained increasing attention~\cite{umutcihanpaper,acr_tufano,acr_automating_code_review_activities_by_large_scale_pre_training,acr_auger_automatically_generating_review_comments} in the last few years. While earlier work often used conventional Machine Learning (ML) and Natural Language Processing (NLP) techniques, recent approaches are increasingly dominated by large-scale pre-trained transformer models and Large Language Models (LLMs) to examine code changes and produce review comments that detect issues and propose potential fixes. The effectiveness of these models largely depends on the quality of their training data. If the data is inadequate or contains errors, language models cannot achieve strong performance. Therefore, reliable and well-curated data is essential.

In code review comment research, existing datasets~\cite{tufano_dataset,turzo_dataset}, though useful, exhibit notable shortcomings. They are often collected from repositories without sufficient curation or preprocessing, leading to noisy data that includes uncivil, irrelevant, or poorly structured comments~\cite{curated_code_reviews}. Such imperfections can impair model training and cause systems to learn misleading or incoherent patterns, ultimately reducing the quality of automated feedback. Therefore, enhancing the cleanliness and reliability of code review datasets is essential to advance automated comment generation and code improvement efforts.

Empirical progress is further hampered by data quality. Widely used public corpora retain significant noise despite heuristic filters. For example, heuristic cleaning still leaves approximately 32\% invalid items~\cite{tufano_dataset}. LLM-based re-analysis finds up to 36\% vague or non-actionable comments that distort training~\cite{zhao2023rightprompts}. Sghaier et al.~\cite{curated_code_reviews} therefore discard about 3.3\% low-relevance items from a 176{,}613-comment corpus to raise data fidelity.

Without (i) a dedicated taxonomy of \emph{smell} patterns and (ii) a \emph{public, curated} benchmark labeled with that taxonomy, researchers cannot fairly compare detectors or study downstream repair. Our study addresses both gaps: we introduce a taxonomy of code review comments and release a dataset of 448 review comments labeled with that taxonomy, mapping coarse ``non-useful'' items from a prior dataset onto explicit smell categories. We additionally examine the labeled benchmark qualitatively to extract and report concrete insights. This resource lays the empirical foundation for systematic detection and, ultimately automated \emph{repair}, of low-quality human review comments.

This study explores the potential of LLMs for analyzing and categorizing code review comments. Complementary to recent LLM-based work on fine-grained review comment classification under an existing taxonomy~\cite{what_makes_a_code_review_useful}, we focus on smell-oriented classification of problematic review comments. LLMs have proven effective in programming and natural language tasks~\cite{survey_evaluation_large_language_models_chang, zan-etal-2023-large-language-models}, particularly through zero-shot learning, where they perform classification without task-specific training data by following prompt instructions. Since labeled data can be scarce, we evaluate both zero-shot and one-shot prompting (one exemplar per category). Based on this setup, we pose the following research questions: 

\noindent
\textbf{RQ1.} How well do LLMs classify review comments into our taxonomy given only the comment text and its linked diff hunk?

\noindent
\textbf{RQ2.} What is the effect of one-shot prompting on classification performance, and which categories are most affected?



\section{Related Work}
\label{sec:related}

\subsection{Code Review Comment Taxonomies}
Prior work on code review comment taxonomies primarily focus on categorizing the function or the intent of the comments. Li et al.~\cite{zhixing} propose a two-level taxonomy of review comments with four first-level groups (\emph{Correctness}, \emph{Decision}, \emph{Management}, \emph{Interaction}) which contain 11 subcategories in total. This taxonomy focuses on the intent of the comment, i.e., the reviewer's goal for which the comment is written. Ochodek et al. ~\cite{journal_classification} introduce a 12-category taxonomy (e.g., code design, style, logic, API, documentation, compatibility, rule definition), derived by analyzing code review comments made on the Wireshark project. This taxonomy focuses on the topic of the comment, i.e., the aspect of code/process that the comment concerns. However, in contrast to our approach, these taxonomies do not make judgments about the quality of the comments.

There are works that include categories capturing quality issues, but the granularity of the issues in these taxonomies are limited. In their investigation of qualities that make code review comments useful, Turzo et al.~\cite{what_makes_a_code_review_useful} introduce a taxonomy which includes \emph{False Positive} as a category. This category captures comments that propose to problems that do not exist. Similarly, while investigating code review automation, Tufano et al. \cite{tufano_dataset} identify certain categories of comments that they discarded from their model evaluations. These include categories such as \emph{Unclear Comment} and \emph{No change asked}. There are also studies that quantitatively classify comments according to quality attributes, though they do not employ a qualitative, taxonomical framework~\cite{EvaCRC, chen_clear_crc}. While notions of problematic code review comments exist, no comprehensive taxonomy, to our knowledge, maps comments to concrete underlying issues.

\subsection{Quality-Focused Code Review Datasets}
\label{sec:noise_datasets}
Previous research on quality-focused code review datasets has largely aimed at annotating comments by specific quality attributes or refining large corpora to reduce noise and improve reliability.

Yang et al.~\cite{EvaCRC} introduce a quality-oriented dataset and an explainable grading scheme for code review comments. They define four context-independent attributes—\emph{emotion}, \emph{question}, \emph{evaluation}, and \emph{suggestion}—and map them to four quality tiers (\emph{poor}, \emph{acceptable}, \emph{good}, \emph{excellent}) to provide reasoned feedback rather than a binary useful/not-useful label. The attributes are derived via triangulation of standards and prior literature, validated on real comments, and operationalized with 16 exemplar mapping rules. In an industrial case study, they annotate 17{,}000 inline comments with an imbalanced tier distribution (\emph{acceptable} 45\%, \emph{good} 38\%, \emph{excellent} 16\%, \emph{poor} 1\%). For automation, their BERT-based multi-label classifier outperforms RF, CNN/RNN, and Transformer baselines, reaching per-attribute F1 of 0.82 (\emph{emotion}), 0.94 (\emph{question}), 0.75 (\emph{evaluation}), and 0.92 (\emph{suggestion}); grade-level macro-F1 is highest for \emph{excellent} (0.79) and lowest for \emph{poor} (0.48). Practitioner interviews also motivate exposing confidence scores and error-correction rules (e.g., limiting the grade under negative emotion) to improve interpretability and trust.

Tufano et al.~\cite{tufano_dataset} explicitly target dataset noise to produce a clean training corpus. They manually annotated 1875 line-linked review comments as relevant or irrelevant using dual annotation with majority resolution, yielding 1676 relevant and 199 irrelevant items. They evaluated n-gram features with Random Forest, J48, and Bayesian Network classifiers in Weka, applying SMOTE for class balancing and feature selection. The best model achieved 91.6\% precision on the relevant class, a modest improvement over the 89\% base rate. They furthermore designed simple keyword heuristics, tuning on 70\% of the labeled set and evaluating on the remaining 30\%; the heuristic filter reached 93.4\% precision for retaining relevant comments and was adopted as a preprocessing step to remove noisy comments from downstream corpora.

Turzo et al.~\cite{turzo_dataset} construct a labeled corpus for OpenDev Nova by performing dual annotation with adjudication, achieving Cohen's Kappa of 0.68. They adopt a taxonomy derived from prior work~\cite{what_makes_a_code_review_useful} with 17 subcategories grouped under five high-level classes: \emph{Functional}, \emph{Refactoring}, \emph{Documentation}, \emph{Discussion}, and \emph{False Positive}. They report a final labeled set of 1{,}828 comments in the paper, while the accompanying replication package contains 1{,}829 labeled comments\footnote{\url{https://github.com/WSU-SEAL/CR-classification-ESEM23}}. Within the labeled set, they report 158 \emph{False Positive} comments. In their scheme, \emph{False Positive} denotes an invalid concern where the raised issue is not actually present, which is a concrete source of dataset noise that must be flagged or filtered when building a clear corpus for downstream repair.

Sghaier et al.~\cite{curated_code_reviews} diagnose noise in a 176{,}613-item, multilingual code-review corpus and use Llama-3.1-70B as an LLM judge to score each comment for relevance, clarity, and conciseness while categorizing type, nature, and civility. Rather than relying on superficial heuristics, they apply a semantics-aware filter: using a relevance threshold of four out of ten, they remove 5{,}895 low-relevance comments ($\approx$3.3\%), yielding a curated set of 170{,}718 items. Examples of removed comments include vague or vacuous remarks such as “Need some edit here?”, “Same here etc :)”, and “This is gross.” Reliability of the automatic judgments is supported by a two-annotator sanity check with high agreement: Cohen’s $\kappa=1.00$ for civility, $0.88$ for type, $0.82$ for nature, and $\kappa=0.85$ (relevance), $0.76$ (conciseness), $0.64$ (clarity). Their study is particularly relevant because it shows that LLM-based curation can target semantic deficiencies in review comments rather than relying only on simple lexical rules.

Liu et al.~\cite{too_noisy_to_learn} target residual noise in open-source code-review datasets and show that heuristic and SVM-based cleaning still leaves many vague or non-actionable comments that degrade training (e.g., CodeReviewer~\cite{acr_automating_code_review_activities_by_large_scale_pre_training}). They argue for semantic, context-aware filtering and motivate it with concrete failures of prior rules. The paper reports that only 64\% of sampled training comments in CodeReviewer are valid; applying LLM classifiers to retain valid items yields precision between 66\% and 85\% and raises the valid ratio in the corpus to about 85\% while reducing the training size by 25--66\%. Models fine-tuned on the cleaned data improve BLEU-4 by 7.5--13\% overall and by 12.4--13.0\% on valid subsets; manual evaluation and large-scale proxies show quality gains up to 24\% in informativeness and 11\% in relevance. The study formalizes ``valid'' versus ``noisy'' review comments for cleaning, documents that earlier heuristics miss many remaining cases, and reports that widely used test sets still contain approximately 32\% noise.

Rahman et al. \cite{predicting_usefulness_of_code_review_comments_rahman} curate an industrial dataset by sampling the 300 most recent inline review comments from each of four commercial systems at ``ABC Company'' via the GitHub API, then discarding PRs created for scaffolding, yielding \(n=1{,}116\) comments. Each comment is labeled as \emph{useful} if a subsequent commit modifies code within 1--10 lines of the comment; the final split is 55.53\% useful vs.\ 44.47\% non-useful. They analyze textual and reviewer-experience features and train RevHelper; a Random Forest reaches approximately \(66\%\) accuracy. This dataset is notable as an early industrial benchmark on comment usefulness, although it is private and not publicly releasable.

Chen et al. \cite{chen_clear_crc} examine the clarity of code-review comments in open-source projects. They perform literature review to generate a first set of potential attributes and refine them via open card sorting to produce a more refined set: relevance, informativeness, and expression. After a preliminary review with 11 industrial practitioners, they design a survey to generate evaluation criteria for each attribute. They manually annotate 2,438 code-diff/comment pairs across nine programming languages according to the generated criteria, achieving strong inter-rater agreement (Cohen’s Kappa = 0.87). They report that roughly 28.8\% of comments fall short on at least one attribute. Among the three attributes, lack of informativeness is the most common deficiency. Based on these findings, the authors propose ClearCRC, a framework for evaluating comment clarity, and benchmark deep learning, machine learning, pre-trained language models, and LLMs on balanced accuracy, precision, recall, and $F_1$ score. They find that pre-trained language models provide encouraging results, reaching a balanced accuracy of $73.04\%$ and $F_1$ score of up to $94.61\%$.

Tufano et al. ~\cite{tufano_code_review_automation}\ provide a comprehensive assessment of current techniques for automating code review across both \emph{code-to-comment} and \emph{code\&comment-to-code} tasks. Their evaluation covers three representative systems spanning neural and IR-based methods and different granularities of code representation. By manually inspecting 2{,}291 predictions from these systems, they derive a fine-grained taxonomy of more than one hundred code-change types and show that existing techniques perform best on relatively simple changes while degrading on more semantically involved ones. A key finding, highly relevant to our work, is that the underlying datasets themselves are far from clean: 574 of the 2{,}291 analyzed instances ($\approx 25\%$) are judged fundamentally problematic and should be discarded for both training and evaluation. They identify recurring issues such as \emph{unclear comment}, \emph{no change asked}, \emph{ignored comment}, and \emph{wrong linking}, and argue that these systematically bias empirical results while motivating better dataset-cleaning pipelines. This directly motivates our focus on a taxonomy and benchmark centered on faults in review comments themselves.

\subsection{Code Review Comment Classification}
Prior research on code review comment classification has primarily focused on categorizing comments along quality dimensions or taxonomies, often emphasizing useful comments. Although our approach classifies comments by smell categories, many of the techniques and attributes remain relevant to our work.


Sarker et al.~\cite{toxic_review} introduce \emph{ToxiCR}, a supervised toxicity detector tailored to software engineering communications and trained on 19{,}651 code review comments from Android, Chromium OS, LibreOffice, and OpenStack, of which 3{,}757 (\(\approx 19.1\%\)) are labeled toxic. Across 10 evaluated algorithms, the best model is BERT with keyword removal, achieving \(A = 95.8\%\) and \(F_{1,\text{toxic}} = 88.9\%\); cross-dataset tests on 4{,}140 Gitter messages yield \(\sim 0.86\) \(F_{1,\text{toxic}}\). They also report a CPU-friendly Random Forest baseline within roughly 1--2 points of the best model. Error analysis highlights pragmatic ``general errors,'' SE polysemy, self-deprecation, and acronym collisions.

Turzo et al.~\cite{turzo_dataset} build a deep, multi-input classifier for code review feedback that fuses CodeBERT embeddings of the surrounding code context (\(\pm 10\) lines) and the comment text with 27 engineered AST/change/file attributes, feeding them through LSTMs and a dense softmax head; labels follow a five-class taxonomy (\emph{Discussion}, \emph{Documentation}, \emph{Refactoring}, \emph{Functional}, \emph{False Positive}) derived from prior work \cite{what_makes_a_code_review_useful}. From OpenDev’s Gerrit, they sample 2,500 comments, obtain dual-annotator labels with \(\kappa = 0.68\) and adjudication, then exclude 672 non--source-file items to yield 1,828 comments. Under 10-fold cross-validation, their best model (CodeBERT for both code and comment) attains 59.3\% accuracy and outperforms a replication of prior work~\cite{fregnan_what_happens_in_my_code_reviews}'s classical ML approach by +18.7 points in accuracy; using BERT for comments is \(\sim 10\) points worse, likely because many comments embed code. Error analysis shows systematic confusion between \emph{Discussion} and \emph{Refactoring} and weak performance on \emph{False Positive} due to class imbalance.

Rahman~\cite{rahman2024} proposes \emph{RefineCode}, a pipeline that first classifies PR review comments as \emph{actionable} versus \emph{non-actionable}, and then assists resolution via similar-review retrieval, Stack Overflow linking, and an LLM chatbot. The study builds an industrial corpus of 9,500 review comments from five private GitHub projects collected in Jan--Nov 2019; 44 developers were invited to label comments and 8 participated (approximately 18\%), yielding 4,313 actionable and 5,187 non-actionable instances. It compares traditional baselines using TF--IDF and sentence embeddings (Sentence-BERT, USE, Mirror-BERT) against fine-tuned transformers. The best detector is BERT with $F_1 = 0.96$ (DistilBERT $F_1 = 0.95$), while strong traditional baselines also reach $F_1 \approx 0.95$. RefineCode then uses retrieval and LLM assistance to propose example-driven fixes.

Nguyen et al.~\cite{exploring_potential_large_language_models_fine_grained_review_comment_classification} study LLM-based classification of review comments under an existing 17-category fine-grained taxonomy~\cite{what_makes_a_code_review_useful} and show that LLMs can outperform a supervised baseline on that task. Complementary to this line of work, our study focuses on a different target: a smell-focused taxonomy and curated benchmark explicitly centered on problematic review comments.

\begin{figure*}[!htbp]
  \centering
  \includegraphics[width=\textwidth]{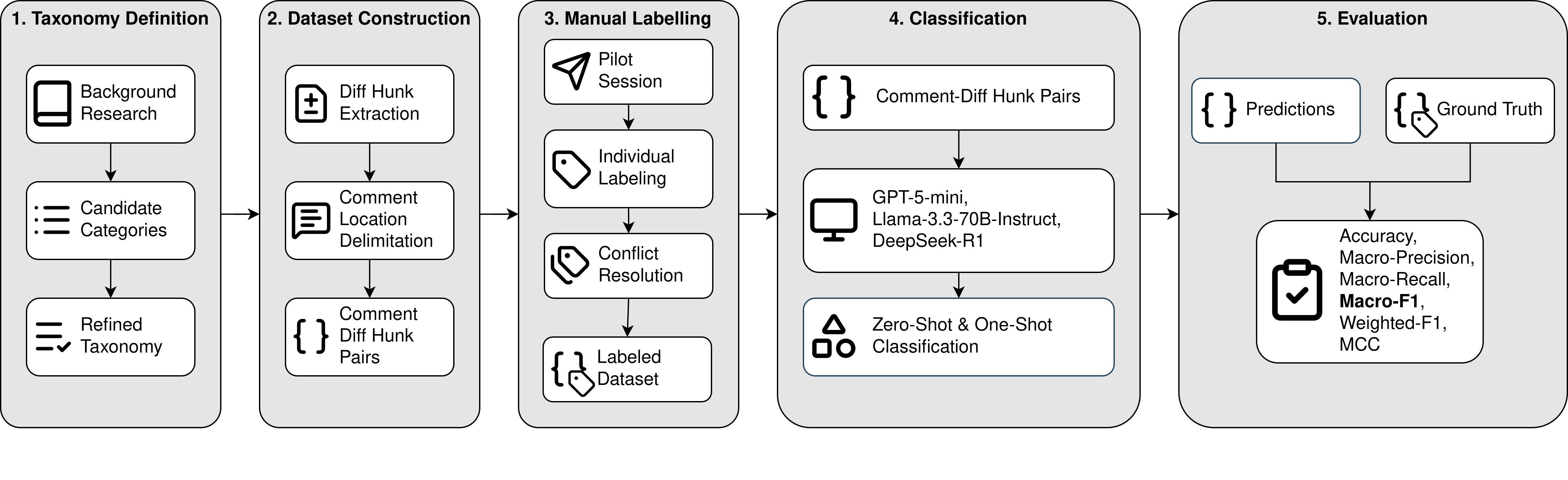}
  \caption{Methodology overview.}
  \label{fig:method}
\end{figure*}

\section{Methodology}
\label{sec:methodology}

This section covers the overall workflow and experimental setup. Figure~\ref{fig:method} provides an overview of our methodology. Section~\ref{sec:taxonomy_definition} covers our smell-focused taxonomy, Section~\ref{sec:methodology_dataset} covers dataset construction and diff extraction, Section~\ref{sec:manual_labelling} covers the manual labeling protocol, Section~\ref{sec:methodology_detection} covers the review comment classification setup, and Section~\ref{sec:methodology_metrics} covers the evaluation metrics.

\begin{table*}[t]
\centering
\small
\begin{tabularx}{\linewidth}{@{}lXXcr@{}}
\toprule
\textbf{Category} & \textbf{Brief Definition} & \textbf{Example Comment} & \textbf{Supporting Studies} & \textbf{Count} \\
\midrule
\textbf{Incorrect} & Claims a specific problem in the code, but that claim is false for the current patch. &
there should be 'True'? (refer to Figure~\ref{fig:incorrect_example}) & \cite{what_makes_a_code_review_useful, turzo_dataset} & 19 \\
\addlinespace[0.4em] \hline \addlinespace[0.4em]

\textbf{Toxic} & Uses hostile, rude, or mocking language instead of a professional tone. &
ugh. this kind of embedded tribal knowledge is just terrible and is an example of why the PCI module is so hard to work with :( & \cite{what_makes_a_code_review_useful, the_stfu_phenomenon, curated_code_reviews, chen_clear_crc, toxic_review} & 10 \\
\addlinespace[0.4em] \hline \addlinespace[0.4em]

\textbf{Unrelated} & Unrelated to the current diff or PR scope. &
Tolkien would be proud & \cite{predicting_usefulness_of_code_review_comments_rahman, curated_code_reviews, chen_clear_crc} & 8 \\
\addlinespace[0.4em] \hline \addlinespace[0.4em]

\textbf{Vague} & Hints at an issue but does not clearly state what/where/why, so it is hard to act on. &
Whoops & \cite{what_makes_a_code_review_useful, curated_code_reviews, chen_clear_crc} & 13 \\
\addlinespace[0.4em] \hline \addlinespace[0.4em]

\textbf{Redundant} & Restates information already obvious from the code, adding no new insight or request. &
weird that this even existed in the first place & \cite{information_needs_in_cr} & 39 \\
\addlinespace[0.4em] \hline \addlinespace[0.4em]

\textbf{Praise} & Primarily praises the code change without suggesting any changes. &
thank you for making this into something somewhat understandable with some code comments. & \cite{bosu2015useful, what_makes_a_code_review_useful} & 70 \\
\addlinespace[0.4em] \hline \addlinespace[0.4em]

\textbf{Useful} & High-value feedback that improves the PR. & & & 289 \\
\hspace{1em} Question & Primarily asks for clarification about this change, without directly requesting a code change. &
Why do you need this mock? & \cite{modern_code_review_bacchelli_intro, what_makes_a_code_review_useful, EvaCRC} & 50 \\
\hspace{1em} Actionable & Explains a concern and recommends a code change. &
you can use decorator instead of this. & \cite{bosu2015useful, curated_code_reviews} & 176 \\
\hspace{1em} Clarification & Adds helpful context/explanation about the code change; does not request a code change. &
I guess this works to fail scheduling because we don't use the PlacementFixture. & \cite{modern_code_review_bacchelli_intro, information_needs_in_cr} & 63 \\
\bottomrule
\end{tabularx}
\caption{Taxonomy of review comments with examples, justifications for each category, and the counts of each category.}
\label{tab:comment_taxonomy}
\end{table*}

\subsection{Taxonomy Definition}
\label{sec:taxonomy_definition}

We define a taxonomy for human review comments that is designed to be 
(i) smell-focused,
(ii) mutually exclusive,
(iii) operational for annotators, and 
(iv) actionable for automated tooling.
Our smell-focused taxonomy additionally includes non-smell categories for useful intents (\emph{Actionable}, \emph{Question}, and \emph{Clarification}), reflecting the mixed nature of real review threads.
As our study focuses on categorizing code review comment smells, 
the taxonomy is derived from and justified by prior empirical studies on code review practices. The full taxonomy is shown in Table~\ref{tab:comment_taxonomy}.

We conducted a focused review of prior work on:
(i) usefulness of code review comments,
(ii) code review comment classification, and
(iii) reviewer intent and communication patterns.
From this review process, we extracted candidate categories.
Following prior research on what makes code review comments useful (e.g., specificity, relevance to the diff, and a constructive tone), we include both \emph{constructive intents} (e.g., actionable suggestions, questions, clarifications) and \emph{low-value} categories that violate these principles (e.g., vague, incorrect, redundant, unrelated, or toxic feedback). Semantically overlapping categories were merged into one, and the names of the categories were modified to have more consistency. We include \textit{Praise} as a smell label. Although praise can support a positive review climate and reinforce desired practices \cite{bosu2015useful}, prior work also finds that praise provides limited actionable value for the current change and is often rated as less useful than other comment types \cite{bosu2015useful,what_makes_a_code_review_useful}.

\subsection{Dataset Construction}
\label{sec:methodology_dataset}
We constructed our dataset by reusing a prior dataset provided by Turzo et al.~\cite{turzo_dataset}. We selected this corpus because it provides per-comment links to the original Gerrit patch set/review thread, allowing direct inspection of the full discussion when additional context was needed, and because its established review-centric taxonomy offers a practical basis that we can re-map into our smell-focused labels.
The dataset (\(N=1,829)\) is composed of code review comments done on the OpenDev Nova project, written in Python, and includes five high-level categories: \emph{Functional}, \emph{Refactoring}, \emph{Documentation}, \emph{Discussion}, and \emph{False Positive}.

\noindent
\textbf{Diff extraction.}
Using the patch set URL provided by the original dataset, we retrieve the corresponding diff and extract the diff hunk associated with each review comment. We exclude comments that are not anchored to any changed lines (i.e., comments not associated with a diff hunk). For retained comments, we also record the span of code referenced by the comment and mark it with delimiters in the hunk.

\noindent
\textbf{Selecting smell candidates.}
For our study, we explicitly target low-quality code review comments. To this end, we extract all comments labeled as \emph{False Positive} or the \textit{Discussion} subcategory \emph{Praise} from this dataset, as these categories are likely to fit the smell categories in our taxonomy. This selection results in a subset of 224 comments.

\noindent
\textbf{Adding useful comments.}
To ensure a more balanced distribution between smell and non-smell comments, we randomly sample an additional 224 comments from the remaining categories in the dataset.
This random sampling results in an equal number of comments for the non-smell set ($N=224$), bringing the total size of our dataset to 448 code review comments.

\subsection{Manual Labeling}
\label{sec:manual_labelling}
We applied a two-stage protocol to assign exactly one category from our taxonomy to each review comment. For each item, annotators were shown (i) the comment text, (ii) the associated unified diff hunk selected for that comment, and (iii) a link to the original review discussion (included in the dataset), which could be used when additional context was needed.
Although a review comment may mention multiple issues, we assign each comment to its primary issue category to maintain a consistent single-label annotation scheme for benchmark construction and evaluation.

\noindent
\textbf{Pilot session.} First, authors A1 (three years of Python experience) and A2 (three years of Python experience) independently labeled a pilot set ($N=10$) drawn uniformly at random from the corpus to calibrate decision boundaries, followed by a short calibration session to refine operational rules. Inter-rater reliability on the pilot was measured using Cohen’s Kappa~\cite{cohen_1960} and was calculated as $\kappa=0.49$. During calibration, we also introduced a dedicated \emph{Clarification} category to capture comments that primarily add explanatory context without proposing a concrete code change.

\noindent
\textbf{Independent labeling and conflict resolution.} After the pilot session, A1 and A2 independently labeled the remaining corpus. Order of the items was randomized per annotator to mitigate order effects, and each item received exactly one label from Table~\ref{tab:comment_taxonomy}. Inter-rater agreement between A1 and A2 before any conflict resolution was $\kappa=0.56$, indicating moderate agreement~\cite{interrater}. The two annotators then conducted a structured reconciliation pass over disagreements, increasing agreement to $\kappa=0.98$ before involving the third author. Remaining unresolved cases were adjudicated in a conflict-resolution meeting, with the third author (A3; 10+ years of empirical software engineering experience) serving as the final arbiter when consensus could not be reached. The final label distribution is summarized in Table~\ref{tab:comment_taxonomy}.

\begin{figure}[t]
  \centering
  \includegraphics[width=\linewidth]{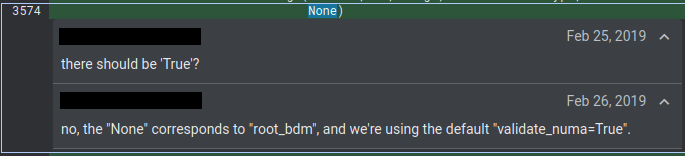}
  \caption{An example of an \emph{Incorrect} review comment, where the reviewer claims a mistake despite the code being correct under the intended logic.}
  \label{fig:incorrect_example}
\end{figure}

\subsection{Review Comment Classification}
\label{sec:methodology_detection}
\textbf{Model selection.}
We use three model families to cover both managed, closed-weights API models and open-weights models: \emph{GPT-5-mini}~\cite{openai_gpt5_mini}, \emph{Llama-3.3-70B-Instruct}~\cite{meta_llama_3.3_70b_instruct}, and \emph{DeepSeek-R1}~\cite{deepseek_r1}. Selection is driven by four constraints: (i) instruction-following reliability under schema-constrained decoding, (ii) code-aware capabilities to parse diffs and identifiers, (iii) cost--latency profile suitable for large batches, and (iv) licensing and reproducibility. \emph{GPT-5-mini} is a compact reasoning-capable model served via a managed API, chosen for robust instruction following and stable structured outputs. \emph{Llama-3.3-70B-Instruct} is an open-weights instruction-tuned model, included as a reproducible baseline with strong general-purpose instruction following and practical performance on code-adjacent text tasks. \emph{DeepSeek-R1} is an open-weights reasoning-focused model, selected for its strong semantic reasoning on code review--related text classification.

\noindent
\textbf{Model Configuration.} For open-weights models, we set the temperature to 0.0 for classification-style prompts to improve output stability and reduce sampling-induced variance. For GPT-5-mini, temperature control is not exposed in the API \cite{openai_gpt5_mini_API}; therefore, we report results under the default sampling behavior while keeping all other prompt and decoding constraints identical (e.g., closed-set, single-label output).

\noindent
\textbf{Prompt design.} We evaluate two inference settings for single-label prediction: \emph{zero-shot} and \emph{one-shot}. Each instance provides (i) the review comment and (ii) its associated unified diff hunk with delimiters marking the span of code marked by the comment. Prompts follow a fixed, delimiter-based layout: (i) a brief task description, (ii) task instructions, (iii) the taxonomy with concise label definitions, (iv) an exemplar block (one-shot only), and (v) the input block containing the comment and diff hunk. Our full prompt templates, alongside our labeled dataset, are available in our replication package\footnote{\url{https://doi.org/10.6084/m9.figshare.31073632}}. In the one-shot setting, we prepend a small set of exemplars drawn from the dataset (and excluded from evaluation), which were manually selected as the clearest representatives for each label and are identical to the examples in \Cref{tab:comment_taxonomy}, to illustrate each label’s decision boundary.
For automatic scoring, we constrain decoding with a simple JSON schema that contains only a label field, and require the model to output only the predicted label, a single string from a fixed set, with no additional fields or free-form text. In case of an improper response according to our schema, we repeat the inference. We use the same base template and context configuration for all three models; the one-shot variant differs only by the additional exemplar block.


\subsection{Evaluation Metrics}
\label{sec:methodology_metrics} 

We evaluate each model using accuracy, macro-precision, macro-recall, macro-F1, weighted-F1, and Matthews Correlation Coefficient (MCC). Because our dataset is highly imbalanced across categories, we use \emph{macro-F1} as the primary metric: it assigns equal weight to each category, preventing majority labels from dominating the evaluation and better reflecting performance on minority labels. We also report MCC as a complementary single-score summary of prediction--label agreement based on the full confusion matrix that remains informative under class imbalance.

\section{Results}
\label{sec:results}
This section reports our classification results and answers the research questions. 

\newcommand{\TableTwoClassificationPerformance}{%
\begin{table}[t]
\centering
\scriptsize
\begin{tabular}{lcccccc}
\toprule
Setting & Acc. & Macro-P & Macro-R & \textbf{Macro-F1} & W-F1 & MCC \\
\midrule
GPT-5-mini, zero-shot & 0.626 & 0.445 & 0.372 & \textbf{0.374} & 0.598 & 0.507 \\
GPT-5-mini, one-shot & 0.645 & 0.491 & 0.408 & \textbf{0.409} & 0.616 & 0.533 \\
LLaMA-3.3, zero-shot & 0.597 & 0.484 & 0.350 & \textbf{0.363} & 0.547 & 0.452 \\
LLaMA-3.3, one-shot & 0.554 & 0.454 & 0.339 & \textbf{0.344} & 0.530 & 0.406 \\
DeepSeek-R1, zero-shot & 0.608 & 0.392 & 0.361 & \textbf{0.360} & 0.593 & 0.486 \\
DeepSeek-R1, one-shot & 0.597 & 0.450 & 0.386 & \textbf{0.388} & 0.591 & 0.485 \\
\bottomrule
\end{tabular}
\caption{Classification performance (Macro-P, Macro-R, and W-F1 refer to Macro-Precision, Macro-Recall, and Weighted-F1 respectively).}
\label{tab:classification_performance}
\end{table}
}

\newcommand{\TableThreePerCategoryPerformance}{%
\begin{table}[!htbp]
\centering
\small
\begin{tabular}{llcccc}
\toprule
Label & Setting & P & R & F1 & Support \\
\midrule
Actionable & GPT-5-mini, zero-shot & 0.667 & 0.777 & 0.718 & 175 \\
 & GPT-5-mini, one-shot & 0.693 & 0.789 & 0.738 & 175 \\
 & LLaMA-3.3, zero-shot & 0.585 & 0.863 & 0.697 & 175 \\
 & LLaMA-3.3, one-shot & 0.606 & 0.749 & 0.670 & 175 \\
 & DeepSeek-R1, zero-shot & 0.668 & 0.760 & 0.711 & 175 \\
 & DeepSeek-R1, one-shot & 0.707 & 0.674 & 0.690 & 175 \\
\addlinespace
Clarification & GPT-5-mini, zero-shot & 0.605 & 0.742 & 0.667 & 62 \\
 & GPT-5-mini, one-shot & 0.561 & 0.742 & 0.639 & 62 \\
 & LLaMA-3.3, zero-shot & 0.562 & 0.581 & 0.571 & 62 \\
 & LLaMA-3.3, one-shot & 0.433 & 0.629 & 0.513 & 62 \\
 & DeepSeek-R1, zero-shot & 0.549 & 0.806 & 0.654 & 62 \\
 & DeepSeek-R1, one-shot & 0.486 & 0.871 & 0.624 & 62 \\
\addlinespace
Incorrect & GPT-5-mini, zero-shot & 0.000 & 0.000 & 0.000 & 18 \\
 & GPT-5-mini, one-shot & 0.000 & 0.000 & 0.000 & 18 \\
 & LLaMA-3.3, zero-shot & 0.000 & 0.000 & 0.000 & 18 \\
 & LLaMA-3.3, one-shot & 0.000 & 0.000 & 0.000 & 18 \\
 & DeepSeek-R1, zero-shot & 0.000 & 0.000 & 0.000 & 18 \\
 & DeepSeek-R1, one-shot & 0.200 & 0.056 & 0.087 & 18 \\
\addlinespace
Praise & GPT-5-mini, zero-shot & 0.896 & 0.870 & 0.882 & 69 \\
 & GPT-5-mini, one-shot & 0.909 & 0.870 & 0.889 & 69 \\
 & LLaMA-3.3, zero-shot & 0.778 & 0.710 & 0.742 & 69 \\
 & LLaMA-3.3, one-shot & 0.814 & 0.696 & 0.750 & 69 \\
 & DeepSeek-R1, zero-shot & 0.902 & 0.797 & 0.846 & 69 \\
 & DeepSeek-R1, one-shot & 0.914 & 0.768 & 0.835 & 69 \\
\addlinespace
Question & GPT-5-mini, zero-shot & 0.549 & 0.571 & 0.560 & 49 \\
 & GPT-5-mini, one-shot & 0.574 & 0.633 & 0.602 & 49 \\
 & LLaMA-3.3, zero-shot & 0.636 & 0.429 & 0.512 & 49 \\
 & LLaMA-3.3, one-shot & 0.714 & 0.408 & 0.519 & 49 \\
 & DeepSeek-R1, zero-shot & 0.792 & 0.388 & 0.521 & 49 \\
 & DeepSeek-R1, one-shot & 0.703 & 0.531 & 0.605 & 49 \\
\addlinespace
Redundant & GPT-5-mini, zero-shot & 0.250 & 0.053 & 0.087 & 38 \\
 & GPT-5-mini, one-shot & 0.600 & 0.079 & 0.140 & 38 \\
 & LLaMA-3.3, zero-shot & 0.000 & 0.000 & 0.000 & 38 \\
 & LLaMA-3.3, one-shot & 0.143 & 0.026 & 0.044 & 38 \\
 & DeepSeek-R1, zero-shot & 0.350 & 0.184 & 0.241 & 38 \\
 & DeepSeek-R1, one-shot & 0.222 & 0.158 & 0.185 & 38 \\
\addlinespace
Toxic & GPT-5-mini, zero-shot & 0.500 & 0.111 & 0.182 & 9 \\
 & GPT-5-mini, one-shot & 0.333 & 0.111 & 0.167 & 9 \\
 & LLaMA-3.3, zero-shot & 1.000 & 0.111 & 0.200 & 9 \\
 & LLaMA-3.3, one-shot & 1.000 & 0.111 & 0.200 & 9 \\
 & DeepSeek-R1, zero-shot & 0.000 & 0.000 & 0.000 & 9 \\
 & DeepSeek-R1, one-shot & 0.500 & 0.111 & 0.182 & 9 \\
\addlinespace
Unrelated & GPT-5-mini, zero-shot & 0.500 & 0.143 & 0.222 & 7 \\
 & GPT-5-mini, one-shot & 0.667 & 0.286 & 0.400 & 7 \\
 & LLaMA-3.3, zero-shot & 0.667 & 0.286 & 0.400 & 7 \\
 & LLaMA-3.3, one-shot & 0.375 & 0.429 & 0.400 & 7 \\
 & DeepSeek-R1, zero-shot & 0.200 & 0.143 & 0.167 & 7 \\
 & DeepSeek-R1, one-shot & 0.250 & 0.143 & 0.182 & 7 \\
\addlinespace
Vague & GPT-5-mini, zero-shot & 0.036 & 0.083 & 0.050 & 12 \\
 & GPT-5-mini, one-shot & 0.077 & 0.167 & 0.105 & 12 \\
 & LLaMA-3.3, zero-shot & 0.125 & 0.167 & 0.143 & 12 \\
 & LLaMA-3.3, one-shot & 0.000 & 0.000 & 0.000 & 12 \\
 & DeepSeek-R1, zero-shot & 0.069 & 0.167 & 0.098 & 12 \\
 & DeepSeek-R1, one-shot & 0.071 & 0.167 & 0.100 & 12 \\
\addlinespace
\bottomrule
\end{tabular}
\caption{Per-category performance.}
\label{tab:per_category_performance}
\end{table}
}

\newcommand{\BestBinaryConfusionMatrix}{%
\begin{table}[!htbp]
\centering
\small
\begin{tabular}{lrr}
\toprule
Gold $\backslash$ Pred & Non-Smell & Smell \\
\midrule
Non-Smell & 262 & 24 \\
Smell & 73 & 80 \\
\bottomrule
\end{tabular}
\caption{Binary confusion matrix for GPT-5-mini, one-shot.}
\label{tab:cm_best_binary}
\end{table}
}

\newcommand{\BestConfusionMatrix}{%
\begin{table*}[!htbp]
\centering
\small
\begin{tabular}{lrrrrrrrrr}
\toprule
Gold $\backslash$ Pred & Actionable & Clarification & Incorrect & Praise & Question & Redundant & Toxic & Unrelated & Vague \\
\midrule
Actionable & 138 & 7 & 0 & 0 & 9 & 2 & 2 & 1 & 16 \\
Clarification & 12 & 46 & 1 & 1 & 2 & 0 & 0 & 0 & 0 \\
Incorrect & 11 & 3 & 0 & 0 & 4 & 0 & 0 & 0 & 0 \\
Praise & 1 & 6 & 0 & 60 & 1 & 0 & 0 & 0 & 1 \\
Question & 16 & 1 & 0 & 1 & 31 & 0 & 0 & 0 & 0 \\
Redundant & 10 & 15 & 0 & 2 & 3 & 3 & 0 & 0 & 5 \\
Toxic & 4 & 0 & 0 & 0 & 2 & 0 & 1 & 0 & 2 \\
Unrelated & 1 & 1 & 0 & 2 & 1 & 0 & 0 & 2 & 0 \\
Vague & 6 & 3 & 0 & 0 & 1 & 0 & 0 & 0 & 2 \\
\bottomrule
\end{tabular}
\caption{Confusion matrix for GPT-5-mini, one-shot.}
\label{tab:cm_best_full}
\end{table*}
}

\noindent
\textbf{RQ1: How well do LLMs classify review comments into our taxonomy given only the comment text and its linked diff hunk?}
Table~\ref{tab:classification_performance} shows that zero-shot performance is moderate (accuracy \(0.597\)–\(0.626\)), but lower under macro-F1 (\(0.360\)–\(0.374\)), reflecting the difficulty of minority and context-dependent categories under label imbalance.
\textit{GPT-5-mini} achieves the strongest zero-shot macro-F1 (\(0.374\)) and MCC (\(0.507\)), while \textit{LLaMA-3.3} and \textit{DeepSeek-R1} are slightly lower in macro-F1 (\(0.363\) and \(0.360\), respectively).

\TableTwoClassificationPerformance

Per-category results in Table~\ref{tab:per_category_performance} reveal a consistent split between linguistically explicit, high-support categories and those that require broader context or verification.
Across models, \textit{Actionable} and \textit{Praise} are reliably identified (e.g., zero-shot F1 \(0.697\)–\(0.718\) for \textit{Actionable}, \(0.742\)–\(0.882\) for \textit{Praise}), and \textit{Clarification} is also comparatively learnable (\(0.551\)–\(0.658\)).
In contrast, \textit{Incorrect} is effectively not detected in zero-shot (F1 \(=0\) for all three models), suggesting that verifying the reviewer’s claim against code semantics is not supported by the limited comment+diff input.
Similarly, \textit{Redundant} and \textit{Vague} remain difficult (zero-shot F1 \(0.000\)–\(0.241\) and \(0.050\)–\(0.143\), respectively), consistent with their dependence on missing thread history (\textit{Redundant}) or underspecification rather than distinctive surface cues (\textit{Vague}).
Overall, zero-shot prompting yields usable performance on frequent operational intents, but it does not yet robustly capture the smell labels that motivate downstream intervention (notably \textit{Incorrect}, \textit{Redundant}, and \textit{Vague}).

\noindent
\textbf{RQ2: What is the effect of one-shot prompting on classification performance, and which categories are most affected?}
One-shot exemplar conditioning may yield small-to-moderate gains in macro-F1, but the effect is model-dependent (refer to Table~\ref{tab:classification_performance}).
For \textit{GPT-5-mini}, one-shot improves macro-F1 from \(0.374\) to \(0.409\) (\(+0.035\)) and increases MCC from \(0.507\) to \(0.533\), producing the best overall setting in our study.
Likewise, \textit{DeepSeek-R1} increases in macro-F1 score (\(+0.028\)) in one-shot setting but shows reduced accuracy and MCC, indicating that exemplar conditioning can trade off global agreement (MCC) for improved minority-class behavior captured by macro-F1.
For \textit{LLaMA-3.3}, both macro-F1 and MCC scores reduce (\(-0.019\) and \(-0.046\), respectively) in one-shot setting.

\noindent
\textbf{Category granularity.} We analyze model performance at the category level (refer to Table~\ref{tab:per_category_performance}) to more accurately judge model behavior.
For \textit{GPT-5-mini}, one-shot increases F1 for \textit{Question} (from 0.560 to 0.602), with smaller gains for \textit{Redundant}, \textit{Unrelated}, and \textit{Vague} that remain low in absolute terms. However, it decreases F1 score for \textit{Clarification} (from 0.667 to 0.639).
For \textit{LLaMA-3.3}, one-shot marginally improves \textit{Question}, \textit{Redundant} and  \textit{Praise}, but it degrades \textit{Vague} to \(0.000\). It also decreases \textit{Actionable} and \textit{Clarification} (\(-0.027\) and \(-0.068\), respectively).
For \textit{DeepSeek-R1}, one-shot improves \textit{Question} and yields non-zero performance on \textit{Incorrect} (\(0.067\)), but it reduces \textit{Actionable} and \textit{Vague}.
Across all models and settings, \textit{Incorrect} remains the dominant gap, reinforcing that this label likely requires verification signals beyond the provided context.
One-shot prompting most consistently improved the models' capability of distinguishing \textit{Question} comments. 

To better understand remaining errors in the best setting, Table~\ref{tab:cm_best_full} shows the confusion matrix for \textit{GPT-5-mini} (one-shot).
A prominent pattern is intent overlap between \textit{Actionable} and \textit{Question}: \(9\) actionable comments are predicted as questions, and \(16\) questions are predicted as actionable, consistent with reviewers phrasing requests as questions.
A second pattern is that \textit{Vague} is used as an uncertainty sink: only \(2/12\) vague comments are recognized, while the model predicts \textit{Vague} \(26\) times in total, primarily absorbing actionable and redundant instances.
Third, \textit{Redundant} is rarely detected (\(3/38\)) and is instead mapped to “useful” labels such as \textit{Clarification} and \textit{Actionable}, which is expected without thread context.
Finally, \textit{Incorrect} is only once correctly predicted (\(1/18\)) and is overwhelmingly interpreted as \textit{Actionable}, suggesting a bias toward treating reviewer claims as valid unless they are explicitly contradicted.

\noindent
\textbf{Smell-NotSmell granularity.} At a coarser granularity, the binary confusion matrix in Table~\ref{tab:cm_best_binary} (NotSmell vs.\ Smell) indicates that the best setting achieves binary F1 \(=0.623\), with comparatively higher precision than recall for the \emph{Smell} class.
This suggests a viable triage behavior (flagging likely smells with fewer false alarms), but with non-trivial miss rate that would still require complementary signals (e.g., retrieved thread context for redundancy and verification-oriented checks for incorrectness).

\TableThreePerCategoryPerformance
\BestConfusionMatrix
\BestBinaryConfusionMatrix

\section{Discussion}
\label{sec:discussion}
This section discusses the implications of our findings. We interpret the results and outline directions for future work.

\begin{figure}[h]
  \centering

  \begin{subfigure}[t]{\linewidth}
    \centering
    \includegraphics[width=\linewidth]{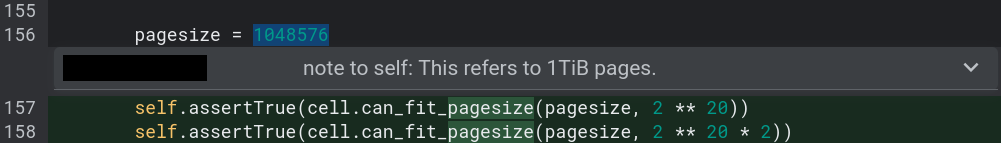}
    \caption{Clarification as code documentation.}
    \label{fig:ex:a}
  \end{subfigure}

  \vspace{0.6em}

  \begin{subfigure}[t]{\linewidth}
    \centering
    \includegraphics[width=\linewidth]{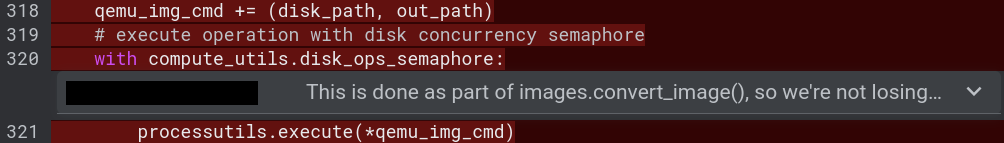}
    \caption{Clarifying why the code was removed.}
    \label{fig:ex:b}
  \end{subfigure}

  \vspace{0.6em}

  \begin{subfigure}[t]{\linewidth}
    \centering
    \includegraphics[width=\linewidth]{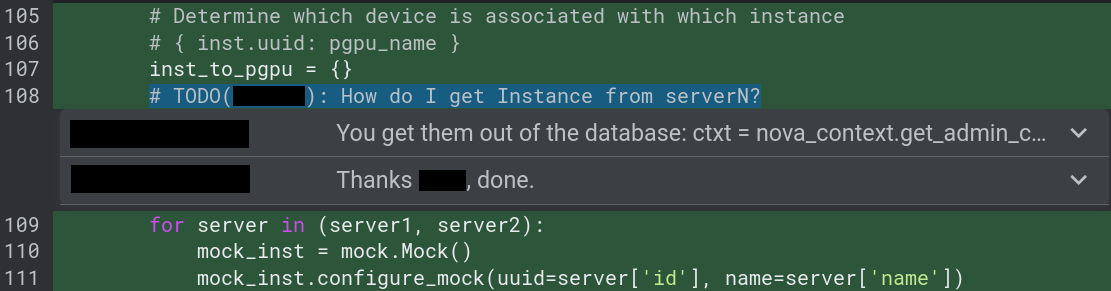}
    \caption{Thread answers an inline TODO.}
    \label{fig:ex:c}
  \end{subfigure}

  \vspace{0.6em}

  \begin{subfigure}[t]{\linewidth}
    \centering
    \includegraphics[width=\linewidth]{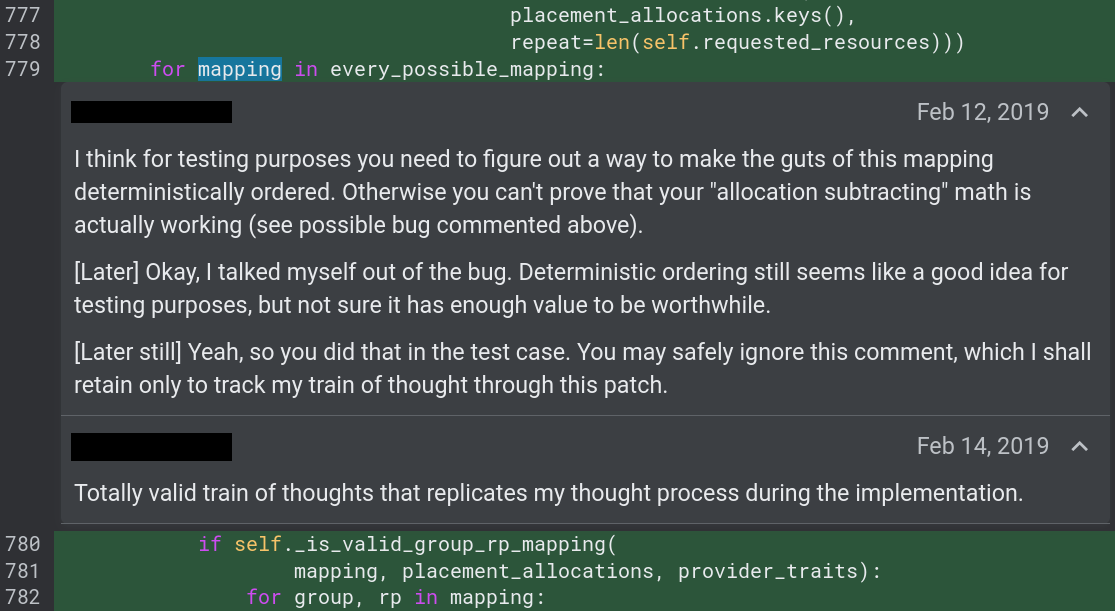}
    \caption{Edited comment revising intent.}
    \label{fig:ex:d}
  \end{subfigure}

  \caption{Example review comment screenshots.}
  \label{fig:ex}
\end{figure}

\subsection{Implications for Researchers}
\label{sec:discussion-researchers}

\noindent\textbf{Treating \emph{Incorrect} comments as context vs.\ as targets.}
Although \emph{Incorrect} review comments can be valuable in real repositories, because they surface and correct misinformation during collaborative development, they are undesirable as \emph{training targets} for automation: a model that learns to produce such comments would be harmful. Therefore, researchers should explicitly separate the role of \emph{Incorrect} comments in the pipeline: they may be retained as \emph{contextual signals} (e.g., to study misconception patterns, measure correction dynamics, or support retrieval that contrasts correct vs.\ incorrect claims), while being excluded (or used as negative exemplar) when training or prompting models that generate review feedback. This distinction encourages dataset designs and evaluations that reflect the dual nature of these artifacts: informative for understanding human review discourse, but misaligned with the goal of producing helpful, correct automated suggestions.

\noindent\textbf{Preferring code-level documentation while recognizing when review-time clarification is necessary.}
In many cases, the same “clarification” does not need to be expressed as a \emph{review comment} at all: it can be captured either as an inline \emph{code comment} (e.g., a brief note explaining a non-obvious choice) or, when possible, by making the code self-explanatory through more meaningful identifier names (see \cref{fig:ex:a}). However, when a reviewer comments on a \emph{deleted} block of code, the rationale for its removal cannot be preserved as a code comment at the referenced location because the referenced code block no longer exists. In such cases, posting a \emph{Clarification} in the review thread is appropriate and often the only viable option (see \cref{fig:ex:b}). For researchers, this motivates separating “should-move-to-code” clarifications (rewrite as code comments or naming improvements) from “change-rationale” clarifications (must remain in the review thread) when constructing datasets and evaluating generation/repair systems.

\noindent\textbf{Resolving context-dependent and referential review comments via thread-level context.}
Some review comments are intrinsically referential—e.g., short acknowledgments like “ditto,” or pointers such as “here” that refer to another review comment posted elsewhere. In these cases, the comment text alone is under-specified, and correct interpretation depends on neighboring comments within the same thread or on cross-comment links to the referenced location. For researchers, this suggests that retrieval and modeling should operate at a broader scope than a single comment–diff pair. Incorporating thread context, temporally adjacent comments by the same reviewer, and explicit cross-references can reduce false \emph{Vague}/\emph{Unrelated} assignments and enable more faithful repair that either expands the context-dependent comment into a self-contained statement or attaches it to its referenced target.

\noindent\textbf{Cross-artifact links between code comments and code review threads.}
Review discussions can "answer" questions embedded in code comments (e.g., a \texttt{TODO} asking how to obtain an instance), after which the author acknowledges completion and the underlying comment arguably becomes obsolete (see \cref{fig:ex:c}). This cross-artifact Q\&A suggests that retrieval for detection/repair should not be limited to the isolated review comment or diff hunk: models may need to jointly consider nearby inline comments, surrounding thread replies, and their temporal ordering to decide whether a code comment should be rewritten, removed, or marked as resolved. For research, this motivates datasets and evaluation setups that explicitly model these links (code-comment -- review-reply), enabling methods that propagate "answered" information back into the codebase and reduce lingering, outdated in-code questions.

\noindent\textbf{Edited review comments and evolving intent.}
Review comments can evolve after posting: reviewers may append later edits that retract an earlier claim, qualify it, or explicitly mark the comment as ignorable while keeping it as a record of their reasoning process (see \cref{fig:ex:d}). This matters for research because the “effective” intent of the comment is not always captured by the first revision: a snapshot taken mid-thread can look \emph{Incorrect} or \emph{Vague}, whereas the final edited version may become a self-correction or meta-note about reasoning.

\noindent\textbf{From smell labels to actionable repair operators.}
Our taxonomy of code review comments provides a structured target space for studying \emph{review comment repair}: each label implies a small, well-scoped set of repairs on the review comment (e.g., \textit{Toxic:}rephrase to remove hostility; \textit{Incorrect:} retract or correct with evidence; \textit{Vague:} specify the affected code and propose a concrete action). This naturally supports modular pipelines that separate (i) detection, (ii) recovery of the reviewer’s underlying intent (issue, location, rationale), and (iii) controlled rewriting under label-specific constraints (e.g., include a pointer to the relevant hunk, avoid context-dependent placeholders such as “here/ditto,” and prevent duplication across comments). A forward-looking direction is to formulate repair as \emph{constrained generation} conditioned on the diff and surrounding thread context, where the model is rewarded for preserving intent while satisfying verifiable constraints induced by the taxonomy. Finally, extending the benchmark with curated “before/after” rewrites (human- or model-authored with validation) would enable both supervised training and fine-grained evaluation of repair quality beyond label accuracy.

\subsection{Implications for Practitioners}
\label{sec:discussion-practitioners}

\noindent\textbf{Adapting triage and responses by comment author (bot vs.\ human).}
In practice, the appropriate handling of a review comment depends not only on its label, but also on \emph{who produced it}. For example, an \textit{Incorrect} comment from an automated bot is often best treated as disposable noise—teams may prefer to delete or ignore it rather than spend effort engaging—whereas an \textit{Incorrect} comment written by a human reviewer typically warrants a corrective clarification to prevent misinformation from persisting in the discussion and to maintain reviewer alignment. Similarly, some categories may be inherently author-dependent: a bot may rarely (or should never) post \textit{Question}-style comments, while humans use questions to negotiate intent and surface missing context. This suggests that deployment should include explicit author-aware policies (e.g., different thresholds, escalation paths, or UI affordances for bot-authored vs.\ human-authored comments) so teams can prioritize effort on interactions that improve shared understanding rather than on correcting machine-generated misfires.

\noindent\textbf{Practical value beyond detection.}
For practitioners, our findings indicate that automated detection of review comment smells can be valuable not only as an inline assistant in daily code review, but also as a teaching and training instrument. Because the taxonomy is phrased in reviewer-friendly terms (\textit{Vague}, \textit{Praise}, \textit{Incorrect}, etc.), a tool based on our detector can surface concrete, actionable feedback to developers about how their comments align with the team’s expectations. One practical extension is to request a brief, evidence-grounded \emph{reason} alongside the predicted label, which could improve transparency and make the feedback easier to learn from.

\noindent\textbf{Onboarding and education.}
One direct application is in onboarding and education. In many teams, junior developers learn “good review practices” informally by reading existing threads and imitating senior reviewers. A detector and repair module integrated into the PR workflow can make this process explicit: when a novice submits a vague or off-topic comment, the system can (privately) flag it and propose a clearer rewrite or an alternative phrasing that better anchors the concern to the diff. Instructors in university software engineering courses or industrial training programs could replay real or synthesized review threads and use our taxonomy to structure exercises where students (i) identify smells in comments, (ii) predict the detector’s label, and (iii) rewrite the comments to be \textit{Useful}. This turns the tool into a formative assessment mechanism rather than merely a gatekeeper.


\noindent\textbf{Reducing review anxiety and encouraging participation.}
A key deployment consideration is to avoid discouraging reviewers from participating. In many teams, the main bottleneck is getting substantive review feedback at all, not perfect wording. Therefore, practitioner-facing tools should present smell labels as optional, supportive suggestions (ideally privately to the author) rather than as enforcement signals, so that reviewers can refine their comments without reducing overall engagement.

\section{Threats to Validity}
\label{sec:threats}

\subsection{Internal Validity}
Our dataset was manually constructed and labeled by the first two authors, with the third author adjudicating disagreements. Despite calibration and shared operational rules, the final labels may still reflect annotator subjectivity. This risk is inherent to our setting because the target of classification is the \emph{intended communicative function} (i.e., what the reviewer meant to accomplish), which is a latent property that cannot be directly observed or measured. Accordingly, our labels should be interpreted as a best-effort approximation of reviewer intent grounded in the available evidence (the comment text and the associated diff hunk), rather than an objective ground truth. To mitigate this threat, we (i) performed a pilot labeling round to refine decision boundaries, (ii) labeled independently before adjudication, and (iii) resolved conflicts through discussion with a final arbiter.

In the one-shot setting, we include exactly one exemplar per label, and these exemplars were manually selected. This introduces a potential source of bias: performance may depend on (i) the decision to use one exemplar per label and (ii) which specific exemplars were chosen (e.g., how prototypical they are, whether they sit near a boundary, or whether they emphasize particular linguistic cues).
To reduce leakage, exemplars drawn from the dataset were excluded from evaluation; however, the selection procedure may still influence results.

During classification, we set the temperature setting to 0.0 for the LLaMA-3.3-70B-Instruct and DeepSeek-R1 models to increase the stability of the performance results. In the case of GPT-5-mini, however, the API did not provide an option to tune the temperature setting. Since the classification results are based on a single execution of a stochastic LLM, observed performance of the model may be affected.

Another threat arises from the possibility that our prompts were not optimal for either model, which could affect output quality, consistency, and intent alignment, and therefore influence measured performance and cross-model comparisons.
While we use the same structured, schema-constrained template across settings to improve comparability, different prompt wordings, context packaging, or added evidence (e.g., broader thread context) might yield different outcomes.
Future work can address this threat via systematic prompt ablations, prompt optimization on a held-out development set, and robustness checks across multiple prompt variants.

\subsection{External Validity}
Our findings may not generalize beyond the specific models and data sources used in this study.
We evaluate detection using three LLMs (\emph{GPT-5-mini}, \emph{Llama-3.3-70B-Instruct}, and \emph{DeepSeek-R1}).
Because LLMs differ in pretraining corpora, instruction tuning, and alignment strategies, the observed effectiveness of zero-shot and one-shot prompting (and the relative strengths across smell categories) may not transfer to other model families or future model versions.

Our benchmark is constructed by filtering and re-labeling instances exclusively from the dataset released by Turzo et al.~\cite{turzo_dataset}. Therefore, our results may partially reflect source-specific characteristics of this single upstream corpus—such as the project context, review culture, primary language, and comment-writing style in Turzo et al.’s dataset—rather than universal properties of code review comments. Because this dataset is publicly available, some models may have seen parts of it during training. If exposure differs across models, cross-model comparisons may be less fair, and performance may appear higher than it would on truly unseen data.
The underlying patches in our corpus are predominantly Python code changes.
As a result, model behavior may partially reflect Python-specific conventions (naming, idioms) and the structure of diffs typical of Python projects.
Future work should replicate the evaluation on additional datasets spanning multiple languages and project contexts, and include a broader set of LLMs to assess robustness.


\subsection{Construct Validity}
A potential threat to construct validity stems from the taxonomy itself, which is an explicit definitional choice for categorizing code review comments. As with any such definition, the taxonomy may not fully or unambiguously capture the underlying conceptual distinctions, potentially affecting the interpretation and validity of the results.

Some review comments may reflect multiple quality issues, so reducing each comment to a single primary category can obscure secondary concerns and blur category boundaries. While annotators were instructed to assign the primary issue reflected in the comment, plausible secondary issues are not represented in the ground truth, which may affect both the interpretation of the label distribution and the measured model performance.



\section{Conclusion}
\label{sec:conclusion}

This paper studies automated classification of human code review comments using LLMs from the comment text paired with its associated unified diff hunk. We benchmark multiple LLMs under a shared schema-constrained, single-label protocol in zero-shot and one-shot settings. Exemplar conditioning is \emph{model-dependent}: it can help some models, but provides limited or negative gains for others.

Our analysis also exposes a key limitation of the comment+diff setting: classification of some labels require evidence beyond the local patch. Verification-oriented and context-dependent judgments remain difficult without thread-level or project-level context, and models may default to uncertainty-oriented labels when evidence is weak.

We make the following contributions: i) a smell-focused taxonomy of review comments, including non-smell intent categories, operationalized for single-label classification; ii) a labeled benchmark of comment--diff pairs (\(N=448\)) for evaluating comment classification with diff hunk context; and iii) a controlled evaluation protocol for comparing zero-shot and one-shot prompting under schema-constrained decoding.


\appendix

\begin{acks}
This study is supported by the Scientific and Technological Research Council of Turkey (TÜBİTAK) under the 2224-A International Scientific Events Participation Support Program.
\end{acks}


\clearpage 

\bibliographystyle{ACM-Reference-Format}
\bibliography{references}

\end{document}